\documentclass[runningheads]{llncs}
\usepackage{graphicx}
\usepackage{amsmath}
\usepackage{ctable}
\usepackage{nicefrac}
\usepackage{booktabs,tabularx}
\usepackage{float}
\usepackage{sidecap}

\usepackage{amssymb}
\usepackage{epstopdf}
\usepackage{subcaption} 
\usepackage[utf8]{inputenc}
\usepackage{authblk}
\usepackage{hyperref}

\makeatletter
\newcommand{\printfnsymbol}[1]{%
  \textsuperscript{\@fnsymbol{#1}}%
}
\makeatother

\begin{document}

% \title{TransVert: Translating Vertebral Annotations from 2D Radiographs to 3D shapes for Spine Modelling(??) (3D Synthesis of Standing Spine from 2D Images? )}
\title{Inferring the 3D Standing Spine Posture \\ from 2D Radiographs}
\titlerunning{Inferring the 3D Standing Spine Posture from 2D Radiographs}

% \author{*}
\author{Amirhossein Bayat\inst{1,2} \thanks{equal contribution} \and
Anjany Sekuboyina\inst{1,2} \printfnsymbol{1} \and 
Johannes C. Paetzold\inst{1} \and
Christian Payer \inst{3} \and
Darko Stern \inst{3} \and
Martin Urschler \inst{4} \and
Jan S. Kirschke\inst{2}\thanks{Joint supervising authors} \and
Bjoern H. Menze\inst{1}\printfnsymbol{2}
}
% index{Bayat, Amirhossein}
% index{Sekuboyina, Anjany}
% index{Paetzold, Johannes C.}
% index{Payer, Christian}
%index{Stern, Darko}
%index{Urschler, Martin}
% index{Kirschke, Jan S.}
% index{Menze, Bjoern}
% \authorrunning{*}
\institute{
Department of Informatics, Technical University of Munich, Germany\\ \and
Department of Neuroradiology, Klinikum rechts der Isar, Germany\\ \and
Institute of Computer Graphics and Vision, Graz University of Technology, Austria\\ \and
School of Computer Science, University of Auckland, New Zealand
\email{amir.bayat@tum.de}\\
}
% \institute{*\\
% \email{amir.bayat@tum.de}}

\maketitle              

\begin{abstract}
The treatment of degenerative spinal disorders requires an understanding of the individual spinal anatomy and curvature in 3D. An upright spinal pose (i.e. standing) under natural weight bearing is crucial for such bio-mechanical analysis. 3D volumetric imaging modalities (e.g. CT and MRI) are performed in patients lying down. On the other hand, radiographs are captured in an upright pose, but result in 2D projections. This work aims to integrate the two realms, i.e. it combines the upright spinal curvature from radiographs with the 3D vertebral shape from CT imaging for synthesizing an upright 3D model of spine, loaded naturally. Specifically, we propose a novel neural network architecture working vertebra-wise, termed \emph{TransVert}, which takes orthogonal 2D radiographs and infers the spine's 3D posture. We validate our architecture on digitally reconstructed radiographs, achieving a 3D reconstruction Dice of $95.52\%$, indicating an almost perfect 2D-to-3D domain translation. Deploying our model on clinical radiographs, we successfully synthesise full-3D, upright, patient-specific spine models for the first time.
\keywords{3D reconstruction  \and Fully Convolutional Neworks \and Spine posture \and Digitally Reconstructed Radiographs}
\end{abstract}

\section{Introduction}

A biomechanical study of spine and its load analysis in upright standing position is an active research topic, especially in cases of spine disorders \cite{biomechanics_review}. Most common approaches for load estimation on the spine either use a general computational model of the spine for all patients or acquire subject-specific models from magnetic resonance imaging (MRI) or computed tomography (CT) \cite{Load_obesity}. While these typical 3D image acquisition schemes capture rich 3D anatomical information, they require the patient to be in a \emph{prone} or \emph{supine} position (lying on one's chest or back), for imaging the spine. But, analysis of the spine's shape and vertebral arrangement needs to be done in a physiologically upright standing position under weight bearing, making 2D plain radiographs a \emph{de facto} choice. A combination of both these worlds is of clinical interest to fully assess the bio-mechanical situation, i.e. to capture patient-specific complex pathological spinal arrangement in a standing position and with 3D information \cite{Load_upright,Load_obesity,load_analysis}. 

In literature, numerous registration-based methods have been proposed for relating 2D radiographs with 3D CT or MR images. In \cite{load_analysis}, the authors propose a rough manual registration of 3D data to 2D sagittal radiographs for the lumbar vertebrae. For the same purpose, in \cite{Gruber}, manual annotations of the vertebral bodies are used as guideline for measuring the vertebral orientations in upright standing position. These methods are time and manual-labour-intensive and thus prone to error. Moreover, both these works use only the sagittal radiographs for vertebra positioning, while ignoring the coronal reformation which is a strong indicator of the spine's natural curvature. Aiming at this objective, \cite{registration_3d} introduced an automatic 3D--2D spine registration algorithm, where the authors propose a multi-stage optimization-based registration method by introducing a metric for comparing a CT projection with a radiograph. However, this metric is hand-crafted, parameter-heavy, and is not learning-based, thus limiting its generalizability. 
%Of special interest is the semi-automatic parametric method introduced in \cite{EOS}. 
In \cite{EOS}, the 3D shape of the spine is reconstructed using a biplanar X-ray device called `EOS'. Hindering its applicability is the high device cost and the lack of its presence in a clinical routine. Recently the problem of reconstructing 3D shapes given 2D images have been explored using deep learning approaches. An approach closest to ours was proposed by Ying et al. \cite{x2ct}, where they introduce a deep neural network to synthesize 3D CT images given orthogonal radiographs using adversarial networks. However, this model is highly memory intensive and fails to synthesise smaller anatomies like vertebrae in 3D. Moreover, it has been evaluated only on digitally reconstructed radiographs (DRR), and its clinical applicability remains to be validated. 

\begin{figure}[t]
\begin{center}
  \includegraphics[width=1.0\linewidth]{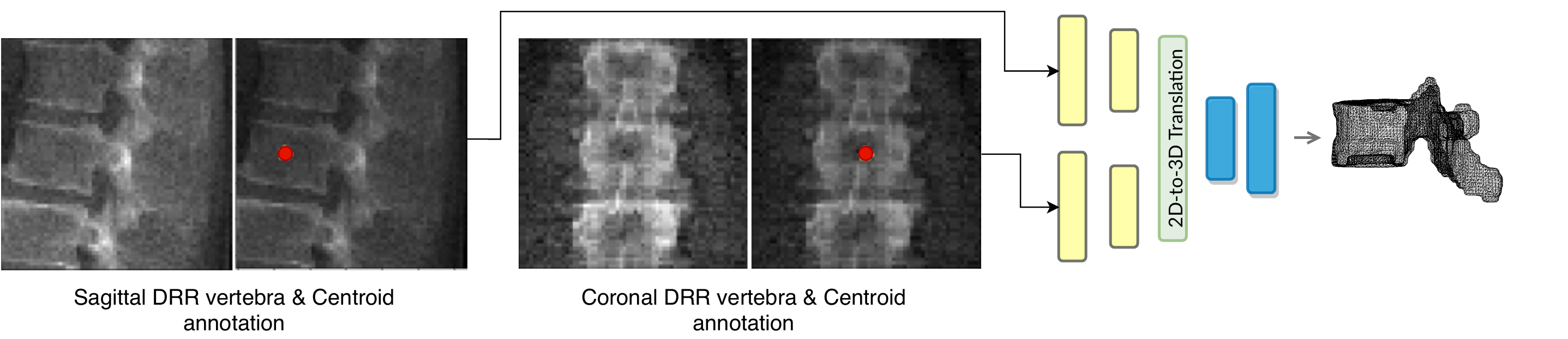}
\end{center}
  \caption{Overview of 2D image to 3D shape translation. The network inputs are 2D orthogonal view vertebrae patches and the centroid indicating the vertebra of interest.}
\label{fig:overview}
\end{figure}

\subsubsection{Motivation} 
The problem of 3D reconstruction of a spine in an anatomically upright position from 2D radiograph images relies on retrieving information from radiographs, which are 2D projections of a 3D object. Spine's sagittal reformation captures crucial information in the form of the vertebral body's and process' shape and its orientation around the sagittal (left-right) axis. However, its orientation around the cranio-caudal and anterior-posterior axes is obfuscated (cf. Fig.~\ref{fig:overview}). 
%This information is available in the coronal reformation, in addition to, for example, the mid-sagittal shape of the vertebra. The coronal reformation also partially visualizes the spinous processes. Therefore, constructing a 3D shape from image data requires an appropriate combination of these orthogonal reformations. Motivated by this, we propose a fully-supervised, computationally efficient, and registration-free approach combining sagittal and coronal 2D radiograph-patches to synthesise the vertebra's 3D shape model. Specifically:
This information is available when combining sagittal with coronal reformations (or lateral with a.p. radiographs). Motivated by this, we propose a fully-supervised, computationally efficient, and registration-free approach combining sagittal and coronal 2D images to synthesise the vertebra's 3D shape model. Specifically:
\begin{itemize}
\item
% We introduce a novel FCN architecture that predicts 3D shapes from 2D orthogonal radiographs.
We introduce a novel fully convolutional network (FCN) architecture for fusing orthogonal radiographs to generate 3D shapes.
% We introduce a novel FCN architecture that combines 2D orthogonal vertebral images and their annotations (e.g vertebral centroids) and maps them to the corresponding 3D shape, as illustrated in Fig.~\ref{fig:overview} 
\item
% Our network is trained on digitally reconstructed radiographs extracted from CT images and supervised by the CT's 3D vertebral masks.
We identify an approach for training the network on synthetically generated radiographs from CT, being supervised by the CT's 3D vertebral masks.
% \item
% Our approach achieves a dice score of \textbf{95.52\%}, indicating a successful inverse mapping of a radiograph, i.e from a two-dimensional to a three-dimensional space.
% % Our approach achieves a mean surface distance of \textbf{xx}mm, indicating a successful inverse mapping from a two-dimensional to a three-dimensional space.
% \item
% Validating its practical applicability, we reconstruct 3D, patient-specific spine models using clinical radiographs.
\item
Validating our approach, we achieve dice score of \textbf{95.52\%} on digitally reconstructed radiographs. We also successfully reconstruct 3D, patient-specific spine models on real clinical radiographs.
\end{itemize}

\section{Methods}
Generating 3D shapes from 2D information is an ill-posed problem. For solving this, we utilize information from two orthogonal radiographs and an annotation on the vertebra of interest while relying on the shape prior learnt by the network.

% \\

%-------------------------------------------------------------------------
\subsection{TransVert: Translating 2D information to 3D shapes}

  The network performing 2D-to-3D synthesis needs to address the following requirements: First, it needs to appropriately combine information in the sagittal and coronal projections to recover 3D information. Second, recovering 3D shapes from 2D projections is inherently an ill-posed problem, requiring incorporation of prior knowledge. Lastly, the size of certain vertebra (towards the scan's periphery) is larger in radiographs compared to their true size due to the cone-beam of gamma-ray source. This effect should be negated when reconstructing the 3D model, i.e. the mapping should not be purely image-based. We address these requirements by proposing the \emph{TransVert} architecture.

\begin{figure}[t]
\begin{center}
  \includegraphics[width=1.0\linewidth]{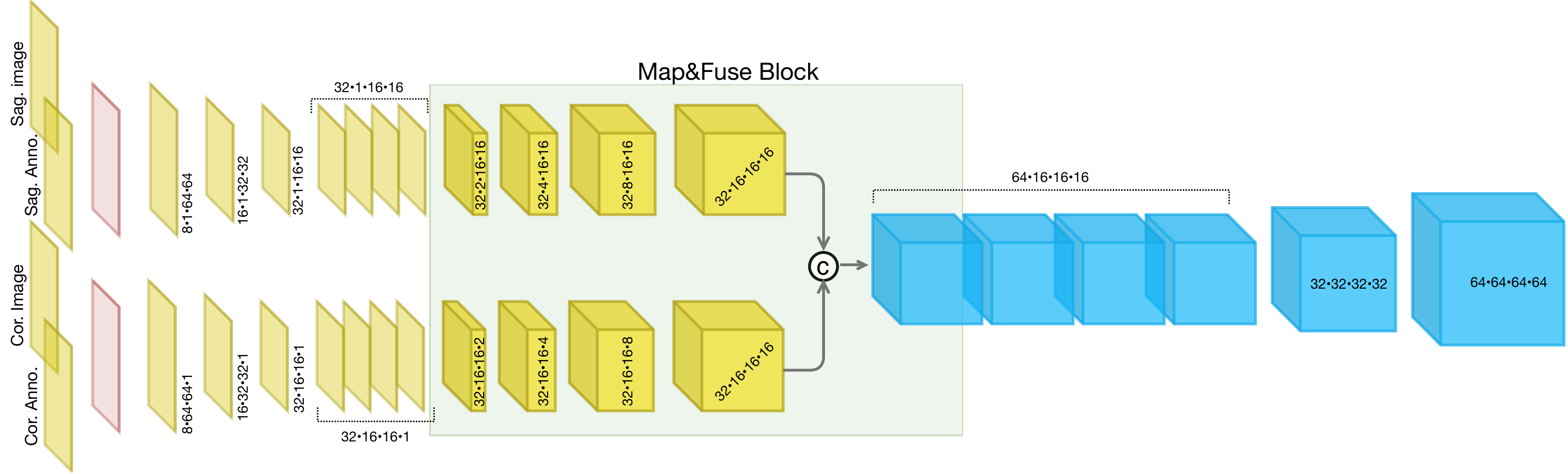}
\end{center}
  \caption{Architecture of \emph{TransVert}. Our model is composed of sagittal and coronal 2D encoders (self-attention module in red), a `map\&fuse' block, and a 3D decoder.}
\label{fig:architecture}
\end{figure}

\subsubsection{Overview} 
TransVert takes four 2D inputs, the sagittal and coronal vertebral image patches and their corresponding annotation images indicating the vertebra-of-interest (VOI). Denoting the 2D vertebral sagittal and coronal reformations by $x_s$ and $x_c$, and their corresponding VOI annotation by $y_s$ and $y_c$, we desire the vertebra's full-body 3D shape, $\mathbf{y}$, as a discrete voxel-map: 

\begin{equation}
\mathbf{y} = G(x_s, x_c, y_s, y_c),
\label{eq:G}
\end{equation}

where $G$ denotes the mapping performed by TransVert. In our case, the VOI-annotation image is obtained by placing a \textbf{disc of radius 1} around the vertebral centroid. In Section \ref{sec:experiments}, we analyze denser annotation choices (vertebral body and full vertebral masks). Ideally, training the TransVert mapping requires radiograph images and their corresponding `real world' 3D spine models. However, this correspondence does not exist and is, in fact, the problem we intend to solve. Thus, TransVert is trained on sagittal and coronal digitally reconstructed radiographs (DRR) constructed from CT images. It is supervised by the corresponding CT images' voxel-level, vertebral segmentation masks. As DRRs are similar in appearance to real radiographs, a DRR-trained TransVert architecture paired with a robust training regime, can be readily deployed on clinical radiographs.    

\subsubsection{Architecture}

TransVert consists of three blocks: a 2D sagittal encoder, a 2D coronal encoder, and a 3D decoder. The three blocks are combined by a `map\&fuse' block. Refer to Fig.~\ref{fig:architecture} for a detailed illustration. The map\&fuse block is responsible for \emph{mapping} 2D representations of each the sagittal and coronal views into intermediate 3D latent representations followed by \emph{fusing} them into a single 3D representation by channel-wise concatenation. This representation is then decoded into a viable 3D voxelized representation by the decoder. Note that the intermediate 3D representation is constructed from orthogonal views. Therefore, map\&fuse block consists of anisotropic convolutions, with an anisotropy along the dimensions that need to be expanded. For example: the anterior-posterior dimension needs to be expanded for a coronal view. Consequently, the convolutional strides and padding directions are orthogonal for each of the view. At the network encoders' input, the vertebral images and VOI-annotations are combined using a self-attention layer. It was empirically observed that the attention mechanism yielded a better performance than a naive fusion by concatenating them as multiple channels.

\subsubsection{Loss} 
Using solely a regression loss leads to converging to a local optimum where a mean (or median) shape is predicted, especially in the highly varying regions of the vertebra such as the vertebral processes. This is rectified by augmenting the loss with an adversarial component which checks the validity of a prediction at a global level. Therefore, TransVert is trained in a fully supervised manner by optimizing a combination of an $\ell_1$ distance-based regression loss and an adversarial loss  based on the least-squared GAN (LSGAN, \cite{LSGAN}). Formally, the TransVert and the Discriminator combination is trained by minimizing the following losses:

% \begin{equation}
% \begin{split}
% \mathcal{L}_{{T}} = & ||\mathbf{y} - \tilde{\mathbf{y}}||_1 + \\
%                     & ({D}_{}(y)-1)^2+({D}_{}(G({x}_{sag}, {x}_{cor})))^2
% \label{eq:1}
% \end{split}
% \end{equation}

\begin{equation}
\mathcal{L}_G =  {\alpha}_G||\mathbf{y} - G(x_s, x_c, y_s, y_c)||_1 + {\alpha}_D\left(D(G(x_s, x_c, y_s, y_c))-1\right)^2~\text{and}
\label{eq:G}
\end{equation}

\begin{equation}
\mathcal{L}_D = (D(\mathbf{y})-1)^2+D(G(x_s, x_c, y_s, y_c))^2,
\label{eq:D}
\end{equation}

% \begin{equation}
% \tilde{\mathbf{y}} =  G({I}_{sag}, {x}_{sag}, {I}_{cor}, {x}_{cor}),
% \label{eq:1}
% \end{equation}

% \begin{equation}
% G({I}_{sag}, {x}_{sag}, {I}_{cor}, {x}_{cor}) = Decoder[Fuse({Encoder}_{sag}({i}_{sag}, {x}_{sag}), {Encoder}_{cor}({i}_{cor}, {x}_{cor}))],
% \label{eq:1}
% \end{equation}

where $D$ represents the discriminator network and $G$ represents the TranVert. ${\alpha}_G$ and ${\alpha}_D$ are weights of loss terms and fixed to ${\alpha}_G$ = 10 and ${\alpha}_D$ = 0.1.
Note that $\mathbf{y}$ is binary valued containing $\{0, i\}$, where $i \in \{8, 9, \dots 24\}$ denotes the vertebral index from T1 to L5. Forcing the network to predict the vertebral index implicitly incorporates an additional prior relating the shape to the vertebral index. Details about the discriminator architecture and the adversarial training regime are provided in the supplemental material. The network is implemented with Pytorch framework on a Quadro P6000 GPU. It is trained till convergence using an Adam optimizer with initial learning rate is 0.0001.

\section{Results}
\label{sec:experiments}
In this section, we describe the creation of DRRs, present an ablative study quantitatively analyzing the contribution of various architectural components, compare various VOI-annotation types, and finally deploy TransVert on real clinical radiographs. 

\subsection{Data} 
Recall that TransVert works with two data modalities: it is trained on DRRs extracted from CT images while being supervised by their corresponding 3D segmentation mask, and it is deployed on clinical radiographs.

% In this section, we describe the simulation of radiograph-like images viz. DRRs, motivate TransVert's architecture. Since the training scheme introduced in this paper is supervised, we need to have pairs of input data and ground truth to train it.  While, TransVert is designed to be deployed on orthogonal radiographs with vertebral annotations to generate 3D model of spine, the 3D shape ground truth for the radiographs is not available. To alleviate this problem, 1) we acquire a spinal CT dataset and segment the spine using a deep neural network. 2) To simulate the real radiographs, we generate Digitally Reconstructed Radiographs (DRR). 3) Finally, we register a spinal atlas with vertebral subregions to the segmentation masks obtained in the step 1. Using these subregions we can extract the vertebral body and vertebral centroids.
% Given the DRRs which resemble the radiographs and annotations as network inputs and the 3D segmentation masks as ground truth, we can train our model TransVert.

\subsubsection{CT data} We work with two datasets: a publicly available dataset for lung nodule detection with 800 chest CT scans \cite{lidc}, and an in-house dataset with 154 CT scans. In all, we work with $\sim12k$ vertebrae split $5:1$ forming the training and validation set, reporting 3-fold cross validated results. Note that very few lumbar vertebrae are visible in \cite{lidc} as it is lung-centred. 
%We perform a $5:1$ split of the data into training and validation set and report the results.
\\
\noindent
\emph{Data Preparation:}
The CT scans are segmented using \cite{segmentation} and the generated masks are validated by an experienced neuro-radiologist in order to consider only accurate ones for the study. These vertebral masks are used for supervision. Generation of the corresponding digitally reconstructed radiographs (DRR) is performed using a ray-casting approach \cite{DRR}, wherein a line is drawn from the radiation source (focal point) to every single pixel on the digitally reconstructed radiographs (DRR) image and the integral of the CT intensities over this line are calculated. Parameters for this generation include the radiation source-to-detector ($=180\text{cm}$ in this work) and the source-to-object distance ($=150\text{cm}$ here). Post the generation of the sagittal and coronal digitally reconstructed radiographs (DRR), patches of size 64$\times$64 are extracted around each vertebral centroid, constituting the image input to TransVert. The second input, viz. the VOI-annotation, can be extracted from the projected segmentation mask.    

% , (similar to our radiograph dataset).  ,  the  TransVert operates \emph{per vertebrae}. We extract a bounding box of size 64$\times$64$\times$64 around each vertebral mask at isotropic 1mm resolution. Also, from sagittal and coronal DRRs we extract 64$\times$64 patches around each vertebral centroid , which constitute the input to the image branch. As input to the annotation branch, we construct three types of annotations from the segmentation mask, one for each of the \textit{V2V}, \textit{B2V}, and \textit{C2V} regimes. We generate full-vertebra annotation for the DRRs by using the same algorithm as in image space on the vertebral mask. For generating the vertebral-body annotation, we register an in-house vertebral atlas (with defined vertebral sub-regions) to the segmentation data CT dataset and isolate only the vertebral body. In all three cases, supervision is provided by the 3D vertebral mask.\\

\subsubsection{Clinical radiographs} We clinically validate TransVert on real long standing radiographs in corresponding lateral and anterior-posterior (a.p.) projections obtained in 30 patients.
%with sagittal and coronal reformations of the same patient. 
Acquisition parameters such as source-to-detector and source-to-object distances were similar to those used for DRR generation. Vertebral centroids needed for the VOI-annotations were automatically generated on both views using \cite{Amir}.
% \cite{Amir}. 

\subsubsection{Data normalization} TransVert is trained on DRRs and tested on clinical radiographs. These data modalities have different intensity ranges, requiring normalization. We observe that z-score normalization works well, i.e.  $\mathcal{I} =  \nicefrac{(\mathcal{I} - \mu_{\mathcal{I}})}{\sigma_{\mathcal{I}}}$, where $\mu_{\mathcal{I}}$ and $\sigma_{\mathcal{I}}$ are the mean and standard deviation of the image $\mathcal{I}$.

\subsection{Experiments}
We perform three sets of experiments validating our proposed approach, aimed at analysing the architectural aspects of TransVert, the data fed into it, and finally its applicability in a clinical setting. Note that a quantitative comparison with the ground truth can be performed only in experiments dealing with DRRs and CT images. Performance evaluation of various settings is compared by computing Dice coefficient and Hausdorff Distance between the predicted 3D vertebral mask and its ground truth from the CT mask.

\subsubsection{Analysing TransVert's architecture} 
\begin{figure}[t]
\begin{center}
  \includegraphics[width=1.0\linewidth]{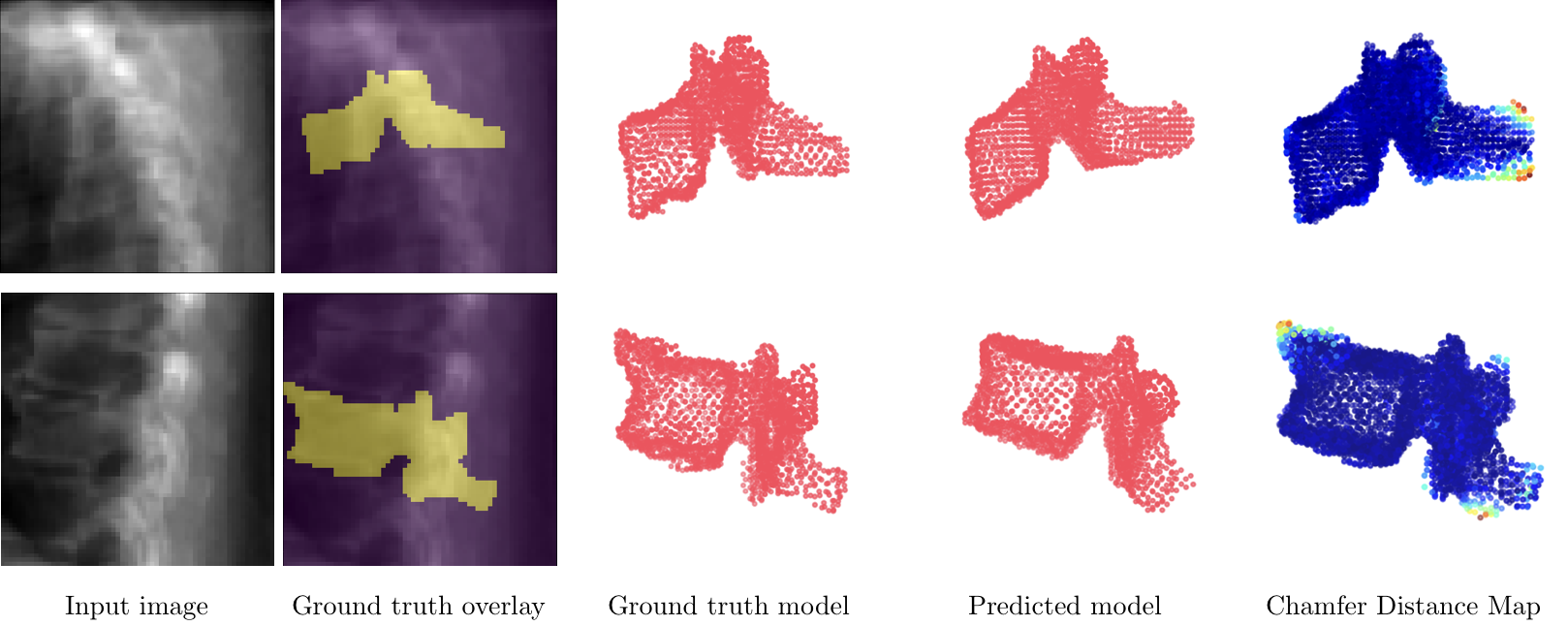}
\end{center}
  \caption{Shape modelling with TransVert on DRRs: First column indicates the image input. Second and third columns visualise the ground truth (GT) vertebral mask and the fourth visualises the predicted 3D shape model. Last column shows an overlayed Chamfer distance map between point clouds of GT and prediction.}
\label{fig:qual_analysis}
\end{figure}

The proposed architecture for TransVert consists of the following architectural choices: fusion of sagittal and coronal views, anisotropic convolutions in the map\&fuse block, a self-attention layer combining the image and the VOI-annotation, and finally, an adversarial component on the loss function. An ablative study over these components is reported in Table \ref{results_tab_architecture}. First, \emph{do we need two views?}. For this, we evaluate the performance of a model that tries to reconstruct 3D shape from only a sagittal image. Next, \emph{do we need anisotropic convolutions?}. For this, we compare two versions of map\&fuse: one with a simple outer product for combining the orthogonal views (Naive View-Fusion) and one with the proposed anisotropic convolutions (TransVert). Observe that a simple fusion of views already outperforms a `sagittal only' reconstruction. Also, anisotropic convolutions outperform fusion of views using outer-products. This can be attributed to the 2D-to-3D learning component involved in the latter. Lastly, \emph{do we need the bells \& whistles on top of TransVert?} Observe that incorporating the self-attention layer in the encoders and an adversarial training regime progressively improved performance, resulting in a Dice of $95.5\%$ and a Hausdorff Distance of 5.11 mm. Fig.~\ref{fig:qual_analysis} illustrates the 3D shape models reconstructed using the proposed architecture. Extracting a point cloud (with 2048 points) from these shapes, we also illustrate a point-wise Chamfer distance map. Observe that a vertebra's posterior region (vertebral process) is hardly visible in the image inputs. Despite this, TransVert is capable of recovering the process, albeit with a certain disagreement between the prediction and ground truth.

% \begin{figure}[t]
% \begin{center}
%   \includegraphics[height=0.3\linewidth]{figures/qual_analysis.png}
% %   \resizebox{\width}{4cm}{\includegraphics[scale=0.4]{figures/qual_analysis.png}}
% \end{center}
%   \caption{Comparison of ground truth to the reconstructed vertebrae. First column is the sagittal image input, second and third columns depict the ground truth (GT) and network prediction respectively. Fourth column shows the chamfer distance between GT and predictions. }
% \label{fig:qual_analysis}
% \end{figure}

\begin{table}[t]
\centering
\setlength{\tabcolsep}{1em}
\small
\renewcommand{\arraystretch}{1.25}
\begin{tabular}{c|c|c}
\specialrule{.1em}{0em}{-.1em}
{Setup} &  Dice (\%) & Hausdorff (mm) \\
\specialrule{.05em}{0em}{-.05em}
Sagittal only & 88.40  & 7.43 \\
Naive View-Fusion (Outer Product) & 92.59  & 6.45 \\
TransVert & 94.75 & 5.75 \\
TransVert + Self Attn. & 95.31 & 5.27 \\
\textbf{TransVert + SelfAttn + Adv.} & 95.52 & 5.11 \\
\specialrule{.1em}{0em}{1em}
\end{tabular}
\caption{Architectural ablative study: The performance progressively improves with addition of each component. (Vertebral centroids are the VOI-annotations here.}
\label{results_tab_architecture}
\end{table}

\subsubsection{Analysing VOI-annotation type} 
Recall that alongside the image input, TransVert requires an auxiliary input indicating the vertebra of interest. We argue that a vertebral centroid suffices. In this study we show that our choice of vertebral centroid performs at a level comparable to a far denser full-vertebra annotation as reported in Table \ref{results_tab_input}. We compare our centoids-to-vertebra (C2V) setup to two other, denser annotations: one where the vertebral body is annotated in the DRR (B2V) and one where the full vertebral body is annotated (V2V).  
% Note that the edges of many vertebral bodies are visible in a radiograph and could be annotated, albeit with considerably greater effort compared to a centroid annotation. 
% We also included the full-vertebra annotation as it is the maximum shape information that can be extracted from a 2D image, despite not being feasible to be annotated in a real radiograph. 
As baseline, we include a setup without any VOI-annotation as an auxiliary input. Note that including the annotation input offers approximately 20\% improvement in the mean Dice coefficient. Observe that a most dense V2V annotations and our C2V annotations perform comparably with only $<1\%$ difference. Therefore, C2V is an obvious choice owing to the ease of marking centroids, more so because of existing automated labelling approaches.

% According to the results, surface of reconstructed vertebra is smoother then the ground truth, since these shapes are conditioned on 2D annotations with no information about the shape surface. Some of these small curves are because of over segmentation, occurred in the segmentation stage. But the model is able to generate similar shapes to the ground truth which preserves the vertebra type and orientation. 
% The qualitative evaluation suggest that the model is able to extrapolate the vertebral process, and also preserves shape and orientation of the vertebrae which are of crucial importance, because the goal is to apply the trained model under C2V method on sagittal and coronal view radiographs patches along wit the vertebral centroid annotations. In radiographs the vertebral centroid is usually annotated, since the vertebral process is not clearly visible.
% An example is demonstrated in Fig.~\ref{img_comparison}. 

% \begin{table}[t]
% \centering
% \setlength{\tabcolsep}{1em}
% \small
% \renewcommand{\arraystretch}{1.25}
% \begin{tabular}{c|c|c}
% \specialrule{.1em}{0em}{-.1em}
% {Input} &  Dice (\%) & Hausdorff (mm) \\
% \specialrule{.05em}{0em}{-.05em}
% No annotation & 76.44  & 14.74 \\
% \textbf{V2V}  & 96.24  & 4.18 \\
% B2V & 95.67  & 4.95 \\
% C2V & 95.31 & 5.27 \\
% \specialrule{.1em}{0em}{1em}
% \end{tabular}
% \caption{Performance of TransVert for various types of input data.}
% \label{results_tab_input}
% \end{table}

\begin{SCtable}
\setlength{\tabcolsep}{1em}
\small
\renewcommand{\arraystretch}{1.25}
\begin{tabular}{c|c|c}
\specialrule{.1em}{0em}{-.1em}
{Input} &  Dice (\%) & Hausdorff (mm) \\
\specialrule{.05em}{0em}{-.05em}
No annotation & 76.44  & 14.74 \\
\textbf{V2V}  & 96.24  & 4.18 \\
B2V & 95.67  & 4.95 \\
C2V & 95.31 & 5.27 \\
\specialrule{.1em}{0em}{1em}
\end{tabular}
\caption{VOI annotation study: Performance drop from a denser (V2V) to a sparser annotation (C2V) is minor, while annotation effort decreases manifold.}
\label{results_tab_input}
\end{SCtable}

% \begin{figure}[H]
\begin{figure}[t]
\begin{center}
  \includegraphics[height=0.75 \linewidth]{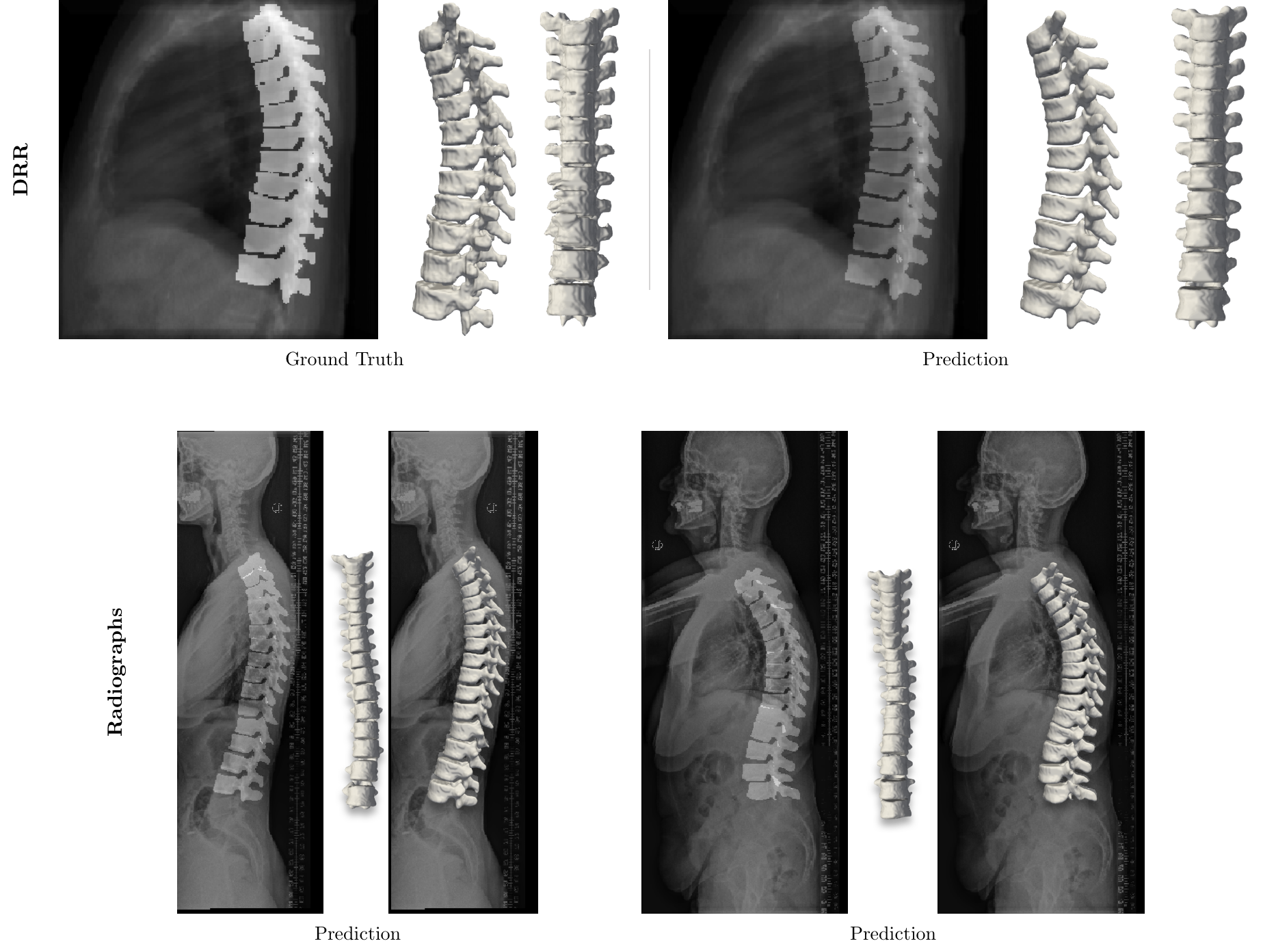}
\end{center}
  \caption{Full 3D spine models: (Top row) Comparison of a DRR-based spine model reconstruction with its CT ground truth mask. (Bottom row) 3D patient-specific spine models constructed from real clinical radiographs. (Best viewed by zooming in.)}
\label{fig:radiograph}
\end{figure}

\subsubsection{2D-to-3D translation in clinical radiographs} 

TransVert works with individual vertebral images and their centroids. A 3D model of the spine can be constructed by stacking the predicted 3D vertebrae models at their corresponding 3D centroid locations. Vertebra's position along the axial and coronal axes is obtained from the sagittal reformation and its sagittal position from the coronal reformation. Fig.~\ref{fig:radiograph} illustrates the results of this process. The top row visualises a 3D spine reconstruction based on 2D DRRs and compares it with the ground truth. More importantly, the bottom row depicts a successful deployment of TransVert in reconstructing the 3D, patient-specific posture of upright standing spine. Note that no 3D ground truth spine model exists for these cases. We visualise the 2D overlay of the segmentation on the radiographs, and the sagittal and coronal view of its 3D shape model, the former overlaid on the radiograph too. Observe that the 3D model's posture matches with that of the radiographs.

\section{Conclusion}
We propose TransVert, a novel architecture trained to infer a full-3D spine model from 2D sagittal and coronal radiographs and sparse centroid annotations. We identify an approach to train TransVert on DRRs in a fully-supervised manner. Along with an ablative study on TransVert's architectural components, we show a successful use case of deploying it on a real-world clinical radiograph. %Pairing TransVert with an automated vertebra labelling algorithm, our setup can be used without manual intervention. 

% As next steps, we intend to   We show that our network is able to generate the 3D shapes given only vertebral centroids as annotations and radiographs. Since these annotations are acquirable automatically, the network can operate in full-automation. 
%The approach we propose could be used to simulate the spinal posture in standing position under the weight bearing, and used in bio-mechanical analysis.

% Acknowledgments---Will not appear in anonymized version
% \vspace{4}
% \midlacknowledgments{Acknowledgements
\subsubsection{Acknowledgements}
This work was funded from the European Research Council (ERC) under the European Union’s Horizon 2020 research and innovation programme (GA637164–iBack–ERC-2014-STG).

\bibliography{midl-samplebibliography}

\end{document}

\title{Supplementary Material}
\author{}
\institute{}
\maketitle              % typeset the header of the contribution
\section{Netowrk Architectures}
% In the following sub-sections, $BN$ denotes Batch normalization, $op$ denotes output padding, $k7$ denotes kernel size of 7$\times$7$\times$7, $ch7$ denotes output channels, $s2$ denotes stride of $[2,2,2]$. For each layer the default is $s1$ and $p0$ if it is not specified.

\begin{table}[]
\centering
\begin{tabular}{ccc}
  \toprule
  \multicolumn{3}{c}{\textbf{Encoders}} \\ 
  ReplicationPad3d(P3) && ReplicationPad3d(P3) \\
  Conv3d(K7Ch8) && Conv3d(K7Ch8) \\
  &Self-Attention \\
  &Conv3d(K7P1S2Ch8) \\
  &Conv3d(K7P1S2Ch16) \\
  &Conv3d(K7P1S2Ch32) \\
  &Residual(Ch32)*4 \\ 

  \multicolumn{3}{c}{\textbf{Map\&Fuse}} \\

  ConvTrans3d(K3P1Ch32OP(1,0,0)S(2,1,1))*4 && ConvTrans3d(K3P1Ch32OP(0,0,1)S(1,1,2))*4\\
  &Concatenate(Ch32,Ch32) \\ 
  \multicolumn{3}{c}{\textbf{Decoder}} \\
  &Residual(Ch64)*4\\
  &ConvTranspose3d(K3P1OP1S2Ch32)\\
  &ConvTranspose3d(K3P1OP1S2Ch16)\\
  &ReplicationPad3d(P3Ch16)\\
  &Conv3d(K7Ch1)
  &\\
  \multicolumn{3}{c}{\emph{Residual Block}} \\
  &ReplicationPad3d(P1Ch\emph{n})\\
  &Conv3d(K3Ch\emph{n}) \\
  &ReplicationPad3d(P1Ch\emph{n})\\
  &Conv3d(K3Ch\emph{n}) \\
  \bottomrule
\end{tabular}
\caption{TransVert architecture.$op$ denotes output padding, $k7$ denotes kernel size of 7$\times$7$\times$7, $ch7$ denotes output channels, $s2$ denotes stride of $[2,2,2]$. For each layer the default is $s1$ and $p0$ if it is not specified. After each layer Batch Normalizatoin and ReLU are used. Double-columns mean two inputs to the subnetwork.}\label{transvert}
\end{table}

\begin{table}[]
\centering
\begin{tabular}{ccc}
  \toprule
  \multicolumn{2}{c}{\textbf{Discriminator}} \\ 
  &SpectralNorm(Conv3d(K4S2Ch64))*4\\
  &SpectralNorm(Conv3d(K4S1Ch64))*4\\
  &Sigmoid() \\ 
  \bottomrule
\end{tabular}
\caption{Discriminator architecture. In both Transvert and discriminator the activation function after layers are ReLU if it is not specified.}\label{discriminator}
\end{table}

% \newpage
\section{More Results}
% We present more qualitative results from different patients in Fig.~\ref{comparison}. The shapes are generated using \textit{body to vertebra} method. Observe the similarity of spinal curve in generated shapes and the ground truth. The model infers the orientation of each vertebra accurately from vertebral body annotations.
% \begin{figure}
% \includegraphics[width=\textwidth]{non_adv_gt_comparison.jpg}
% \caption{Generating 3D spine shapes given 2D vertebral body annotations. The generated volume (left) using \textit{body to vertebra} method is compared to the ground truth (right) for 3 patients. Once trained, the model is able to infer the vertebral shape and orientation from 2D annotations.} \label{comparison}
% \end{figure}

\begin{figure}[H]
\begin{center}
  \includegraphics[height=0.9\linewidth]{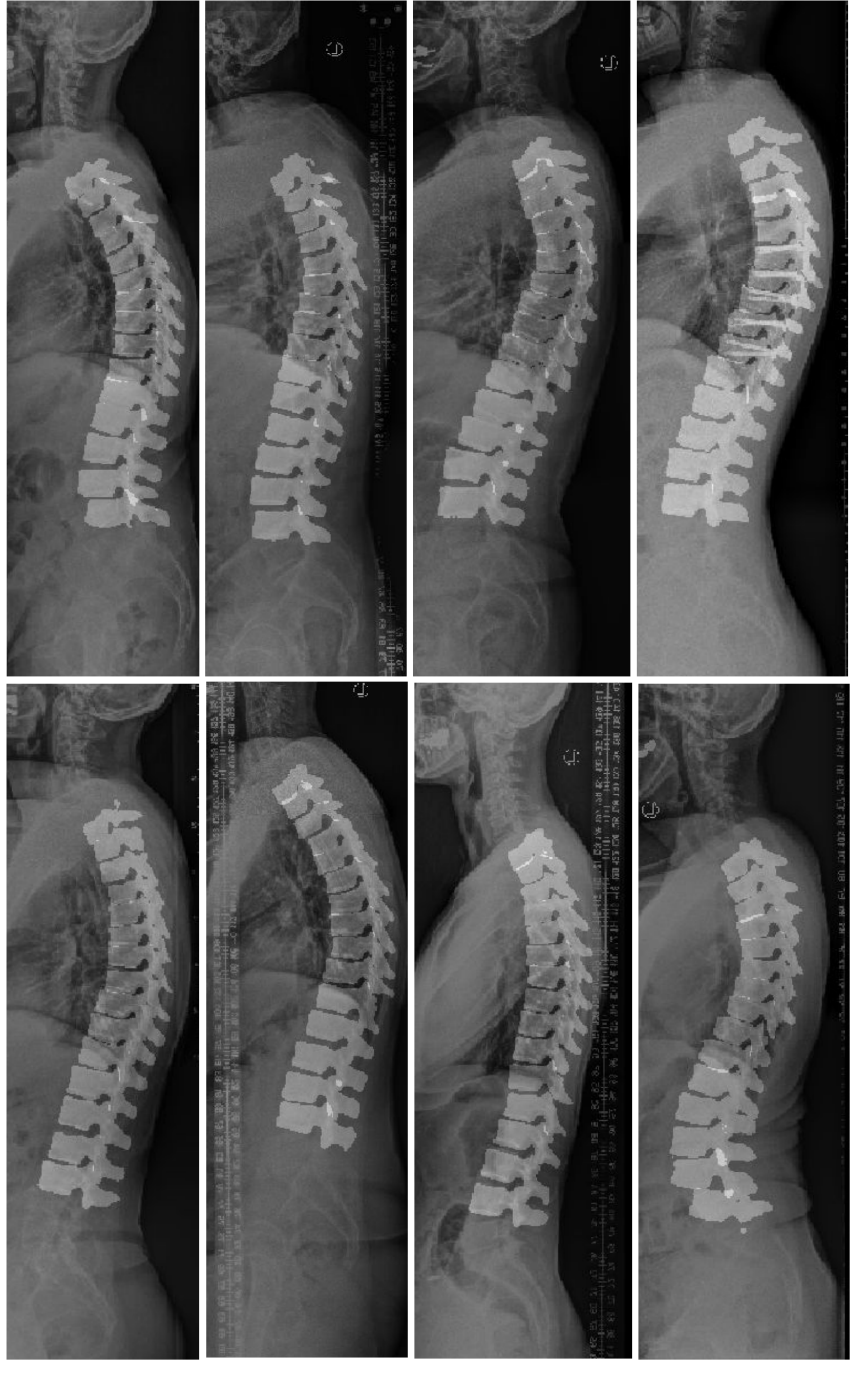}
\end{center}
  \caption{3D standing spine postures overlaid on the clinical radiographs. Observe that the inferred postures match the spinal curves in the radiographs. The right-most patient in top row has implants in spine, but the model is robust against these artifacts in the image.}
\label{fig:radiograph}
\end{figure}

% \begin{figure}
% \includegraphics[width=\textwidth]{nonadv_gan_gt.jpg}
% \caption{From left to right, the results of the \textit{body to vertebra} and \textit{Unsupervised body to vertebra} methods and the ground truth.} \label{comparison}
% \end{figure}